\documentclass[pra,twocolumn,showpacs]{revtex4}
\usepackage{times}
\newcommand{\tr}{\mbox{Tr}}
\newcommand{\rk}{\mbox{Rank}}

\begin{document}
\title{
    Positive Maps Which Are Not Completely Positive
   }
\author{Sixia Yu}
\affiliation{
    The Institute of Theoretical Physics,
    Academia Sinica, Beijing 100080, P.R. China
    }
\date{\today}
\begin{abstract}
The concept of the {\em half density matrix} is proposed. It
unifies the quantum states which are described by density matrices
and physical processes which are described by completely positive
maps. With the help of the half-density-matrix representation of
Hermitian linear map, we show that every positive map which is not
completely positive is a {\em difference} of two completely
positive maps. A necessary and sufficient condition for a positive
map which is not completely positive is also presented, which is
illustrated by some examples.
\end{abstract}
\maketitle

Entanglement has become one of the central concept in quantum
mechanics, specially in quantum information. A quantum state of a
bipartite system is {\em entangled} if it cannot be prepared
locally or it cannot be expressed as a convex combination of
direct product states of two subsystems. This kind of state is
also called {\em inseparable}. Though easily defined, it is very
hard to recognize the inseparability of a mixed state of a
bipartite system.

An operational-friendly criterion of separability was proposed by
Peres \cite{peres}. This criterion is based on the observation
that the partial transposition of a separable density matrix
remains positive semidefinite. That the partial transposition of a
density matrix is {\em not} positive semidefinite infers the
inseparability of the density matrix. This provides a necessary
condition for the separability. There exist entangled states with
positive partial transposition, which exhibit bound entanglement~\cite{bound}. Examples of such kind were first provided in
Ref.~\cite{phoro} and then constructed in Ref.~\cite{upb}
systematically with the help of unextendible product basis.

Later on, by noticing that the transposition is a {\em positive map} (to
be described later in details), a necessary and sufficient condition
of the separability was proposed in Ref.~\cite{horox}: A bipartite
state is separable iff it is still positive semidefinite under all
positive maps acting on a subsystem. In other words, a density
matrix of a bipartite system is inseparable iff there exists a
positive map acting on a subsystem such that the image of the
density matrix is not positive semidefinite. Hence the
inseparability can be recognized by positive maps which are not
{\em completely positive}.

Completely positive maps, which are able to describe the most
general physical process \cite{kraus}, are better understood than
positive maps which are not completely positive. Positive maps
from Hilbert space ${\cal H}_2$ (two-dimensional) to ${\cal H}_2$
or ${\cal H}_3$ are all decomposable~\cite{decp}, which are
characterized by transposition and completely positive maps only.
As a result in the cases of ${\cal H}_2\times{\cal H}_2$ and
${\cal H}_2\times {\cal H}_3$ the transposition criterion is also
a sufficient condition for separability \cite{horox}. Therefore
further understandings of positive maps which are not completely
positive will facilitate the recognition and classification of the
inseparable mixed states.

As a direct calculation will show, under an orthonormal and
complete basis $\{|n\rangle\}_{n=0}^{L-1}$ the transposition of an
$L\times L$ matrix $\rho$ can be expressed as
\begin{equation}\label{tsp}
\rho^{T}=\tr\rho-\sum_{m,n=0}^{L-1}\;\sigma_{mn}\;\rho\;\sigma_{mn}^\dagger,
\end{equation}
where $\sigma_{mn}=(|m\rangle\langle n|-|n\rangle\langle
m|)/\sqrt2$. We see immediately that the transposition is a
difference of two completely positive maps. And this statement
will be proved to hold true for all positive maps which are not
completely positive, which will be also characterized by a
necessary and sufficient condition in this Letter.

For this purpose we shall first develop an extremely useful tool
--- {\em half density matrix} that unifies the description of the
quantum states and physical processes. And then we derive a
half-density-matrix representation of an arbitrary Hermitian
linear map from which our main results are obtained. Along with
the introduction of the concept of half density matrix, its
relations to the ensembles and the purifications of mixed states
are clarified and its applications in the field of quantum
information such as quantum teleportation \cite{tele} are also
presented.

Normally, quantum states, pure or mixed, are described by density
matrices, positive semidefinite operators (whose eigenvalues are
all nonnegative) on the Hilbert space of the system. Because of
its property of positive semidefinite the density matrix $\rho$
can always be written as $\rho=TT^\dagger$ where matrix $T$ is
called here as the {\em half density matrix} (HDM) for a quantum
state.

Obviously, the half density matrix for a given density matrix is
not unique. For example $TU$ and $T$ are corresponding to the same
mixed state $\rho=TT^\dagger$ whenever $U$ is unitary. Generally,
the half density matrix $T$ for a mixed state $\rho$ of an
$s$-level system is an $s\times L$ rectangular matrix with $L\ge
r=\rk(\rho)$, i.e., a linear map from an $L$-dimensional Hilbert
space ${\cal H}_L$ to an $s$-dimensional Hilbert space ${\cal
H}_s$. The rank $r$ of the density matrix equals to the rank of
the half density matrix $T$ and $r=1$ for pure state.

Under an orthonormal and complete bases
$\{|m\rangle\}_{m=0}^{s-1}$ and $\{|n\rangle\}_{n=0}^{L-1}$ of
Hilbert spaces ${\cal H}_s$ and ${\cal H}_L$, a typical half
density matrix of dimension $s\times L$ can be constructed as
$T_e=V^\dagger(\Delta_s,0_{s\times(L-s)})$ , where $\Delta_s$ is a
diagonal $s\times s$ matrix formed by all the square roots of the
eigenvalues of $\rho$ (the singular numbers of $T_e$) and $V$ is
an $s\times s$ unitary matrix diagonalizing the density matrix
$\rho$. Obviously we have $\rho=T_eT_e^\dagger$. As a direct
result of the {\em singular number decomposition} of an arbitrary
matrix \cite{sglr} we have the following

{\bf\em Lemma:} Given a density matrix $\rho$ of an $s$-level
system, an ${s\times L}$ matrix $T$ is a half density matrix for
$\rho$, i.e., $\rho=TT^\dagger$, if and only if there exists an
$L\times L$ unitary matrix $U$ such that $T=T_eU$.

When written explicitly in the established basis, the relation
$\rho=T_eT_e^\dagger$ results in exactly an ensemble formed by all
the eigenvectors $V^\dagger|m\rangle$ of the mixed state, which is
referred to as {\em eigen-ensemble} here. In this way every half
density matrix $T$ of a mixed state $\rho$ corresponds to an
ensemble of the mixed state. The above {\em Lemma} tells us that
every ensemble of a given mixed state is related to the
eigen-ensemble by a unitary matrix which has been proved by other
means \cite{hjw}. Therefore the half density matrix of a density
matrix is physically equivalent to an ensemble of the
corresponding mixed state.

Every mixed state $\rho$ admits a {\em purification} \cite{schu},
a pure state $|\phi\rangle$ of a bipartite system including this
system as a subsystem such that
$\rho=\tr_2|\phi\rangle\langle\phi|$. Under the established basis,
a general pure state in ${\cal H}_s\otimes{\cal H}_L$ is
\begin{equation}
|\phi\rangle=\sum_{m=0}^{s-1}\sum_{n=0}^{L-1}
C_{mn}|m\rangle_1|n\rangle_2 :=T|\Phi_L\rangle.
\end{equation}
Here pure state
$|\Phi_L\rangle=\sum_{n=0}^{L-1}|n\rangle_1|n\rangle_2$ lives in
Hilbert space ${\cal H}_L\otimes{\cal H}_L$  and $T$ is a linear
map from ${\cal H}_L$ to ${\cal H}_s$ acting on the first
$L$-dimensional Hilbert space ${\cal H}_L$. Under the given bases
linear map $T$ is represented by an $s\times L$ rectangular matrix
with matrix elements given by $\langle m|T|n\rangle=C_{mn}$. When
the pure state $|\phi\rangle$ is normalized we have $\tr(T^\dagger
T)=1$. Alternatively, we also have
$|\phi\rangle=T^T|\Phi_s\rangle$ with state $|\Phi_s\rangle$
defined in ${\cal H}_s\otimes{\cal H}_s$ similar to state
$|\Phi_L\rangle$. The linear map $T^T:{\cal H}_s\mapsto {\cal
H}_L$ acts on the second ${\cal H}_s$ and it is represented by the
transposition of $T$ under the established basis.

Tracing out the second system we obtain a reduced density matrix
of the first subsystem $\rho_s=TT^\dagger$ and similarly
$\rho_L=T^{T}T^{*}$ for the second subsystem. That is to say $T$
is the HDM for the reduced half density matrix
$\rho_s=\tr_L|\phi\rangle\langle\phi|$ of the first subsystem and
its transposition $T^T$ for
$\rho_L=\tr_s|\phi\rangle\langle\phi|$. Thus a one-to-one
correspondence between a normalized pure state $|\phi\rangle$ of a
bipartite system, a purification, and a linear map $T$ satisfying
$\tr(T^\dagger T)=1$, a half density matrix, is established.
Therefore a half density matrix $T$ is also equivalent to a
purification of the mixed state. The linear map $T$ is also
referred to as the half density matrix of a bipartite pure state,
which is unique by definition. If $s=L$ the polar decomposition of
$T$ will result in the useful Schmit-decomposition.

The pure bipartite state is separable iff the rank of its half
density matrix is one. For a pure product state
$|v\rangle_s|w\rangle_L$ the half density matrix  is
$|v\rangle\langle w^*|$ where $|w^*\rangle$ is the {\em index
state} of state $|w\rangle$ defined by $|w^*\rangle=\langle
w|\Phi_L\rangle$ \cite{schu}. For later use we define a {\em
mirror operator} $M_L=|\Phi_L\rangle\langle\Phi_L|$ in the Hilbert
space ${\cal H}_L\otimes{\cal H}_L$, which has the property
$\langle w^*|M_L|w^*\rangle=|w\rangle\langle w|$. The partial
transposition of the mirror operator $X=M_L^{T_1}$ is in fact the
exchanging (or swapping) operator introduced by
Werner~\cite{werner} (denoted as $V$ there).

As an application, we consider a state
$|\phi\rangle_{12}|\psi\rangle_3$ of a tripartite system with all
three subsystems 1,2 and 3 being $s$-level systems. Let $T_\phi$
denote the HDM of the bipartite state $|\phi\rangle_{12}$ and
$|k;l\rangle_{23}=T_{kl}|\Phi_s\rangle_{23}$ denote an orthonormal
complete basis for systems 2 and 3 with HDMs $T_{kl}$ satisfying
$
 {\rm Tr}T_{kl}T^\dagger_{k^\prime l^\prime}
 =\delta_{kk^\prime}\delta_{ll^\prime}
$
for orthogonality and $\sum_{kl}T_{kl}{\cal O}T^\dagger_{kl}=\tr
{\cal O}$ for completeness. We then have expansion
\begin{equation}
|\phi\rangle_{12}|\psi\rangle_3=\sum_{k,l=0}^{s-1}T_\phi\:
T_{kl}^*\;|\psi\rangle_1|k;l\rangle_{23}.
\end{equation}
This describes exactly a quantum teleportation of an unknown
quantum state $|\psi\rangle$ from system 3 to system 1 when both
$T_\phi$ and $T_{kl}$ are unitary or state $|\phi\rangle_{12}$ and
basis $|k;l\rangle_{23}$ are maximally entangled states \cite{ys}.

The mixed state $\rho_{sL}$ of an ($s\times L$) bipartite system
can also be equivalently and conveniently characterized by HDMs of
pure bipartite states. Let $\{|\phi_i\rangle,p_i\}_{i=1}^{R}$ be
an ensembles of $\rho_{sL}$ we have
\begin{equation}\label{hd}
\rho_{sL}=\sum_{i=1}^Rp_i|\phi_i\rangle\langle
\phi_i|=\sum_{i=1}^RA_iM_LA_i^\dagger,
\end{equation}
where we have denoted $A_i$ as the half density matrix of the pure
state $\sqrt{p_i}|\phi_i\rangle$, i.e.,
$\sqrt{p_i}|\phi_i\rangle=A_i|\Phi_L\rangle$. Obviously HDMs
defined by $\tilde A_i=\sum_jU_{ij}A_j$ characterize the same
density matrix whenever $U$ is a unitary matrix. And from the {\em
Lemma} we know that given a density matrix this is the only
freedom that the half density matrices can have.

The density matrix expressed in the form as in Eq.(\ref{hd}) can
be easily manipulated by local operations. For example the density
matrix under operation $U_s\otimes U_L^*$ is transformed to
density matrix specified by half density matrices
$U_sA_iU_L^\dagger$. The tilde operation $\rho\mapsto \tilde\rho$
introduced in Ref.~\cite{2qubit} to obtain explicitly the
entanglement of formation of two-qubit is simply an anti-linear
transformation $\tilde A_i=\tr A_i^\dagger-A_i^\dagger$.

In the discussions above we have defined the half density matrices
for the states of a single system, for pure bipartite states, and
for mixed bipartite states. The physical processes can also be
characterized by half density matrices. A general physical process
which can include unitary evolutions, tracing out one system, and
general measurements is described by {\em trace-preserving
completely positive maps} \cite{kraus,schu}, which is a special
kind of Hermitian linear map.

A Hermitian linear map sends linearly Hermitian operators to
Hermitian operators that may live in different Hilbert spaces. Let
${\cal L}$ denote a general Hermitian linear map from Hilbert
space ${\cal H}_L$ to ${\cal H}_s$. Because the map ${\cal L}$ is
linear the map ${\cal L}\otimes {\cal I}_L$ is also a Hermitian
linear map from ${\cal H}_L\otimes{\cal H}_L$ to ${\cal
H}_s\otimes{\cal H}_L$, where ${\cal I}_L$ denotes the identity
map on ${\cal H}_L$. Recalling that the mirror operator
$M_L=|\Phi_L\rangle\langle\Phi_L|$ is defined on ${\cal
H}_L\otimes{\cal H}_L$, its image
\begin{equation}\label{rel3}
H_{sL}={\cal L}\otimes {\cal I}_L(M_L)
\end{equation}
is therefore a Hermitian operator in ${\cal H}_s\otimes{\cal
H}_L$. Let $|\psi^+_i\rangle=A_i|\Phi_L\rangle$ $(i\le i_+)$
denote the eigenvectors corresponding to the positive eigenvalues
of $H_{sL}$ and $|\psi^-_i\rangle=B_i|\Phi_L\rangle$ $(i\le i_-)$
to negative eigenvalues of $H_{sL}$, where $i_\pm$ is the number
of the positive/negative eigenvalues of $H_{sL}$.  We then have
\begin{equation}
H_{sL}=\sum_{i=1}^{i_+}A_iM_LA_i^\dagger-
       \sum_{i=1}^{i_-}B_iM_LB_i^\dagger,
\end{equation}
in which the norms of the eigenvectors $|\psi^\pm_i\rangle$ have
been taken to be the absolute values of corresponding eigenvalues.
Because the eigenvectors corresponding to different eigenvalues
are orthonormal we have $\tr(A_iB^\dagger_j)=0$ for all $i$ and
$j$. In this sense two families of half density matrices $\{A_i\}$
and $\{B_i\}$ are {\em orthogonal} to each other.

For a pure state $P_w=|w\rangle\langle w|$ in the Hilbert space
${\cal H}_L$ we have $P_w=\langle w^*|M_{L}|w^*\rangle$ where
$|w^*\rangle$ is the index state of $|w\rangle$. As a result we
have ${\cal L}(P_w)=\langle w^*|H_{sL}|w^*\rangle$. Taking into
account of the linearity of the Hermitian map ${\cal L}$, we
finally obtain
\begin{equation}\label{hlm}
{\cal L}(H)=\sum_{i=1}^{i_+}A_iHA_i^\dagger-
               \sum_{i=1}^{i_-}B_iHB_i^\dagger,
\end{equation}
where $H$ is an arbitrary Hermitian matrix in ${\cal H}_L$. This
is called the {\em half-density-matrix representation} of a
Hermitian linear map. As one result we have
\begin{equation}\label{id}
\langle\Phi_s|{\cal I}_s\otimes{\cal L}(\Sigma_{sL})|\Phi_s\rangle
=\tr(H_{sL}\Sigma_{sL}^T)
\end{equation}
for an arbitrary Hermitian matrix $\Sigma_{sL}$ in ${\cal
H}_s\otimes{\cal H}_L$. As another consequence, a one-to-one
correspondence between the Hermitian maps ${\cal L}:{\cal
H}_L\mapsto{\cal H}_s$ and Hermitian matrix $H_{sL}$ (an
observable) in ${\cal H}_s\otimes{\cal H}_L$ can be established
\begin{equation}\label{rel2}
{\cal L}(H)=\tr_L(H_{sL}H^T)
\end{equation}
in addition to Eq.(\ref{rel3}).

The HDM representation of Hermitian linear map is not unique.
Suppose two integers $M\ge i_+$ and $N\ge i_-$ and let $SU(M,N)$
denote the pseudo-unitary group formed by $(M+N)\times(M+N)$
matrices satisfying $S\eta S^\dagger=\eta$ where $\eta=I_M\oplus
(-I_N)$ and $I_{M(N)}$ is the $M\times M$ ($N\times N$) identity
matrix. If we define a family of HDMs $\{T_i\}_{i=1}^{M+N}$ as
$T_i=A_i$ $(1\le i\le i_+)$, $T_i=B_i$ $(M+1\le i\le M+i_-)$ and
$T_i=0$ otherwise and take an arbitrary element $S$ of $SU(M,N)$,
a new family of HDMs $\{\tilde T_i\}_{i=1}^{M+N}$ defined by
$\tilde T_i=\sum_jS_{ij}T_j$ represents the same Hermitian linear
map
\begin{equation}\label{rel}
{\cal L}(H)=\sum_{i=1}^{M}\tilde T_iH\tilde
T^\dagger_i-\sum_{j=1}^{N}\tilde T_jH\tilde T^\dagger_j.
\end{equation}

A positive map is a special Hermitian linear map which maps any
positive semidefinite operator to a positive semidefinite
operator. A Hermitian linear map ${\cal S}:{\cal H}_L\mapsto{\cal
H}_s$ is positive if and only if $\tr({\cal
S}(Q_L)P_s)=\tr(H_{sL}^TP_s\otimes Q_L)\ge0$ for all pure product
state $P_s\otimes Q_L$ where $H_{sL}={\cal S}\otimes{\cal
I}_L(M_L)$. In the following ${\cal S}$ is always a positive map.

A completely positive (CP) map is a positive map which keeps its
positivity when the system it acts on is embedded as a subsystem
in an arbitrary larger system. That is, for a CP map ${\cal
S}:{\cal H}_L\mapsto{\cal H}_s$ and an arbitrary positive integer
$k$ the induced map ${\cal S}\otimes{\cal I}_k$ from ${\cal
H}_L\otimes{\cal H}_k$ to ${\cal H}_s\otimes{\cal H}_k$ is
positive.

However it is enough to check whether the image $H_{sL}={\cal
S}\otimes {\cal I}_L(M_L)$ of the mirror operator $M_L$ is
positive semidefinite or not. If it is positive semidefinite, then
the negative part in the HDM representation Eq.(\ref{hlm})
disappears, which yields exactly the {\em operator-sum
representation} of a CP map~\cite{schu}
\begin{equation}\label{osr}
{\cal S}(\rho)= \sum_{i=1}^{i_+}A_i \rho A^\dagger_i.
\end{equation}
If the trace is preserved, we have further $\sum_iA_i^\dagger
A_i=1$. Therefore the operator-sum representation of a CP map can
also be referred to as a {\em half-density-matrix representation}.
Especially, if $H_{sL}$ equals to the identity matrix $I_s\otimes
I_{L}$, the corresponding CP map is simply the trace operation
${\cal S}_T(\rho)=I_s\tr\rho$.

A positive map which is not completely positive (non-CP) is
nonetheless a Hermitian map so that it has a HDM representation as
Eq.(\ref{hlm}), from which we obtain ${\cal S}={\cal S}_A-{\cal
S}_B$ where two CP maps ${\cal S}_{A,B}$ are represented by HDMs
$\{A_i\}$ and $\{B_i\}$ respectively. Two CP maps ${\cal S}_{A,B}$
are said to be orthogonal if their HDMs are orthogonal to each
other, i.e., $\tr(A_iB^\dagger_j)=0$ for all $i,j$. We see that
$H_{sL}$ can not be positive semidefinite.

Conversely, if the Hermitian matrix $H_{sL}$ has at least one
negative eigenvalue then it determines a non-CP positive map. Let
$|\psi\rangle$ denote an eigenvector corresponding to one of the
negative eigenvalues of $H_{sL}$ and
$P_\psi=|\psi\rangle\langle\psi|$. From identity (\ref{id}) we see
immediately that ${\cal I}_s\otimes{\cal S}(P^T_\psi)$ is not
positive semidefinite, i.e., map ${\cal S}$ is not completely
positive. We note that the eigenspace corresponding to the
negative eigenvalues of $H_{sL}$ contains no product state because
of positivity. To summarize, we have the following

{\bf\em Theorem:} Every positive map which is not completely
positive is a difference of two orthogonal completely positive
maps; A Hermitian linear map ${\cal S}:{\cal H}_L\mapsto{\cal
H}_s$ is positive but not completely positive if and only if for
all pure product state $P_s\otimes Q_L$ in ${\cal H}_L\otimes{\cal
H}_s$ we have $\tr(H_{sL}P_s\otimes Q_L)\ge0$ while $H_{sL}={\cal
S}\otimes{\cal I}_L(M_L)$ is {\em not} positive semidefinite.

This theorem provides an obvious way to construct a non-CP
positive map form ${\cal H}_L$ to ${\cal H}_s$. First, we choose a
proper Hermitian matrix $H_{sL}$ in ${\cal H}_s\otimes{\cal H}_L$
satisfying the conditions specified in the above theorem. Then a
non-CP positive map ${\cal S}:{\cal H}_L\mapsto{\cal H}_s$ is
determined by ${\cal S}(\rho_L)=\tr_L(H_{sL}\rho_L^T)$.

As the first example we consider the the exchanging operator
defined in ${\cal H}_L\otimes{\cal H}_L$ by $X=M_L^{T_1}$ or
explicitly
\begin{equation}\label{X}
X=\sum_{m,n=0}^{L-1}|m,n\rangle\langle n,m|.
\end{equation}
The exchanging operator $X$ has two eigenvalues $\pm1$ and
$\sigma_{mn}|\Phi_L\rangle$ $(m>n)$ are the eigenvectors
corresponding to eigenvalue $-1$. Therefore $X$ is not positive
semidefinite and for any pure product states
$|pp\rangle=|v\rangle|w\rangle$ we have $\langle
pp|X|pp\rangle=|\langle v|w\rangle|^2\ge 0$ as specified by the
above theorem. In fact the resulting non-CP positive map on ${\cal
H}_L$ is exactly the transposition $\rho^T=\tr_2(X\rho^T)$. By
writing $X$ in its diagonal form we obtain $\rho^T={\cal S}_T
(\rho)-{\cal S}_\sigma(\rho)$, where the CP map ${\cal S}_\sigma$
is represented by HDMs $\{\sigma_{mn}\}$ and ${\cal S}_T$ is the
trace operation.

As the second example we consider a Hermitian matrix in ${\cal
H}_L\otimes{\cal H}_L$ defined by $H_R=I_{L}\otimes I_L-M_L$. It
is not positive semidefinite because
$\langle\Phi_L|H_R|\Phi_L\rangle<0$ and for every product states
$|pp\rangle$ we have $\langle pp|H_R|pp\rangle= 1-|\langle
v|w^*\rangle|^2\ge0$. Accordingly, a non-CP positive map is
defined on ${\cal H}_L$ as $\Lambda(\rho)=\tr\rho-\rho$, which
provides the {\em reduction criterion} \cite{reduct,red0}: Every
inseparable state in ${\cal H}_L\otimes{\cal H}_L$ which loses its
positivity under map ${\cal I}_L\otimes\Lambda$ is distillable and
in the distillation procedure provided in Ref.~\cite{reduct} the
HDM of pure bipartite state serves as the filtering operation.
Because $\Lambda(\rho)={\cal S}_\sigma(\rho^T)$, the reduction map
$\Lambda$ is a decomposable positive map, which is generally of
form ${\cal S}_d(\rho)={\cal S}_{1}(\rho)+{\cal S}_{2}(\rho^T)$
with ${\cal S}_{1,2}$ being two CP maps.

The last example makes use of an unextendible product basis
\cite{upb}, a set of orthonormal product basis
$\{|\alpha_i\rangle|\beta_i\rangle\}_{i=1}^{S}$ of ${\cal
H}_s\otimes{\cal H}_L$ where $S<sL$ and there is no other pure
product state that is orthogonal to this set of basis. If we
denote
$P=\sum|\alpha_i\rangle\langle\alpha_i|\otimes|\beta_i\rangle\langle\beta_i|$
then $\tilde\rho=(1-P)/(sL-S)$ represents an inseparable states
with positive partial transposition. If we define
$$\epsilon=\min_{|\alpha\rangle|\beta\rangle}\langle\alpha|\langle\beta|P|\alpha\rangle
|\beta\rangle$$ it can be sure that $0<\epsilon\le S/sL$
\cite{bt}. Denoting $\rho_0$ as a normalized density matrix in
${\cal H}_s\otimes{\cal H}_L$ which has the property
$\tr(\rho_0\tilde\rho)> 0$, we define a Hermitian matrix as
$H_\epsilon=P-\epsilon d\rho_0$ where
$$\frac1d=\max_{|\alpha\rangle|\beta\rangle}\langle\alpha|\langle\beta|\rho_0|\alpha\rangle
|\beta\rangle$$ and $1\le d\le sL$. Matrix $H_\epsilon$ is not
positive semidefinite since $\tr H_\epsilon\tilde\rho=-\epsilon
d\tr(\rho_0\tilde\rho)<0$ and for an arbitrary pure product state
$\langle\alpha|\langle\beta|H_\epsilon|\alpha\rangle|\beta\rangle\ge0$.
If we choose $\rho_0=I_{s}\otimes I_L/sL$ then a non-CP positive
map is defined by
\begin{equation}
{\cal S}_\epsilon(\rho)=\sum_{i=1}^ST_i\rho
T_i^\dagger-\epsilon\tr\rho,
\end{equation}
where $T_i=|\alpha_i\rangle\langle\beta_i^*|$ is the half density
matrix of the product base $|\alpha_i\rangle|\beta_i\rangle$. In
Ref.~\cite{bt} $\rho_0$ is taken as a maximally entangle state and
$d=\min(s,L)$. Positive map ${\cal S}_\epsilon$ is indecomposable
because ${\cal I}_s\otimes{\cal S}_\epsilon(\tilde\rho)$ is not
positive semidefinite while ${\cal I}_s\otimes{\cal
S}_d(\tilde\rho)$ is positive semidefinite for any decomposable
map.

In conclusion, the concept of the half density matrix was studied
and its applications to the quantum information are discussed in
some detail. Based on the half-density-matrix representation of a
Hermitian linear map, we proved that every positive map which is
not completely positive is a difference of two completely positive
maps. A necessary and sufficient condition for a non-CP positive
map is given, which provides a way of constructing such kind of
maps. Some examples are also presented. Further applications of
the half density matrix in the quantum information and other
fields can be expected and the understandings of positive maps
provided here may be helpful the recognition of the inseparable
quantum states and to the quantification of the
entanglement~\cite{moe}.

The author gratefully acknowledges the financial support of K. C.
Wong Education Foundation, Hong Kong.

%\begin{references}

\end{document}